\documentclass[twocolumn,showkeys,aps,prb,showpacs]{revtex4-1}
\usepackage{graphicx}
\usepackage[CJKbookmarks,dvipdfm,colorlinks,linkcolor=blue,citecolor=blue]{hyperref}

\begin{document}

\title{ Piezoelectric  properties of  $\mathrm{Ga_2O_3}$: a first-principle study}

\author{San-Dong Guo and Hui-Min Du}
\affiliation{School of Electronic Engineering, Xi'an University of Posts and Telecommunications, Xi'an 710121, China}
\begin{abstract}
The compounds exhibit piezoelectricity, which demands to   break inversion symmetry,  and then to be a semiconductor.
For  $\mathrm{Ga_2O_3}$, the orthorhombic case ($\epsilon$-$\mathrm{Ga_2O_3}$) of common five phases  breaks inversion symmetry. Here, the piezoelectric tensor  of $\epsilon$-$\mathrm{Ga_2O_3}$ is reported by using density functional perturbation theory (DFPT). To confirm semiconducting properties of $\epsilon$-$\mathrm{Ga_2O_3}$, its electronic structures are studied by using generalized gradient
approximation (GGA) and   Tran and Blaha's modified Becke and Johnson (mBJ) exchange potential. The gap value of 4.66 eV is predicted with mBJ method, along with the the effective mass tensor for electrons at
the conduction band minimum (CBM) [about 0.24 $m_0$]. The mBJ gap is very close to the available experimental value. The elastic   tensor $C_{ij}$ and piezoelectric stress tensor $e_{ij}$ are attained by DFPT, and then piezoelectric strain  tensor $d_{ij}$ are calculated from $C_{ij}$ and $e_{ij}$. In this process, average mechanical properties of $\epsilon$-$\mathrm{Ga_2O_3}$ are estimated, such as
bulk modulus, Shear modulus, Young's modulus and so on. The calculated $d_{ij}$  are comparable and even higher than commonly used piezoelectric materials such as $\alpha$-quartz, ZnO, AlN  and GaN.

\end{abstract}

\pacs{71.20.-b, 77.65.-j,  77.65.Bn ~~~~~~~~~~~~~~~~~~~~~~~~~~~~~~~~~~~Email:sandongyuwang@163.com}

\maketitle

\section{Introduction}
Wide-band gap semiconductors have potential application  in high-power electronics, which requires high
frequency, temperature and power. Gallium oxide ( $\mathrm{Ga_2O_3}$) has received a lot of attention
as  a wide band gap transparent semiconducting oxide\cite{g0,g1,g2,g3,g4,g5}. The  $\mathrm{Ga_2O_3}$ has  five different phases,
commonly referred to as $\alpha$, $\beta$,  $\gamma$, $\delta$ and  $\epsilon$, the monoclinic $\beta$ phase of which is the most thermodynamically
stable with the energy gap  4.6-4.9 eV, transparency up to the UV-C range, and very high breakdown voltage\cite{g0,g2}.
Piezoelectric materials can  convert mechanical energy
to electrical energy, which  have  potential application in sensors and energy harvesting\cite{g6}.
The ZnO, GaN  and InN semiconductors with non-centrosymmetric
wurtzite-structure  are wildly used in the piezotronic and
piezo-phototronic devices, and  their nanostructures   have potential applications in  electromechanical
coupled sensors and nanoscale energy conversion\cite{g7,g8,g9,g10}.

 For piezoelectric materials,
inversion symmetry need be eliminated. The bravais lattice, space group, point group and inversion center  of five different polymorphs of  $\mathrm{Ga_2O_3}$ are shown in \autoref{tab0-gr}. It is clearly seen that $\epsilon$-$\mathrm{Ga_2O_3}$  breaks inversion symmetry
and hence can exhibit piezoelectricity. The $\epsilon$-phase of $\mathrm{Ga_2O_3}$ is confirmed
as the second most stable structure after $\beta$-$\mathrm{Ga_2O_3}$\cite{g11}.  When $\epsilon$-$\mathrm{Ga_2O_3}$ is epitaxially stabilized,  the symmetry will prevent the transform back into  $\beta$-phase.
The electronic structures of $\epsilon$-$\mathrm{Ga_2O_3}$ have been reported, and the predicted gap is 2.465 eV with GGA\cite{g12}, 2.32 eV with PBEsol, 4.62 eV  with  B3LYP\cite{g13} and  4.26 eV with HSE\cite{g14}.  The experimental gap is 4.41 eV by angle-resolved photoemission spectroscopy (ARPES) experiments\cite{g14}, and is 4.6 eV from photoconductivity and optical absorption\cite{g15}.
The $\epsilon$-$\mathrm{Ga_2O_3}$ is predicted to have a large spontaneous polarization (0.23-0.26  $\mathrm{C/m^2}$)\cite{g11,g13},  along with  piezoelectric coefficient $e_{33}$=0.77 $\mathrm{C/m^2}$\cite{g11}. Recently, piezoelectric strain constants ($d_{ij}$) of
$\epsilon$-$\mathrm{Ga_2O_3}$ are calculated from piezoelectric stress constants ($e_{ij}$) and elastic constants ($C_{ij}$)\cite{g16}. The  $e_{ij}$ are attained by polarization-strain relation, and the $C_{ij}$ are calculated by energy-stain relation\cite{g16}. Here, we use DFTP to attain the $C_{ij}$ and $e_{ij}$, and then calculate the $d_{ij}$ by $e$ matrix multiplying  $C$ matrix inversion.  To ensure the reliability of our results, the piezoelectric  properties of commonly used piezoelectric materials such as  ZnO, AlN  and GaN are also studied by DFTP, and make a comparison with the related experiments\cite{zno,aln,aln-1,aln-2,aln-3}.  The mBJ is used to study the electronic
structures of $\epsilon$-$\mathrm{Ga_2O_3}$, and the calculated mBJ gap 4.66 eV is very close to experimental values\cite{g14,g15}. The mBJ is as cheap as local density approximation (LDA) or GGA, thus can be used  to study very large systems  such as doping $\epsilon$-$\mathrm{Ga_2O_3}$ in an efficient way.

\begin{table}
\centering \caption{The bravais lattice, space group, point group and inversion center  of five different polymorphs of  $\mathrm{Ga_2O_3}$. }\label{tab0-gr}
  \begin{tabular*}{0.49\textwidth}{@{\extracolsep{\fill}}cccccc}
  \hline\hline
  Name   &$\alpha$ & $\beta$   & $\gamma$& $\delta$&$\epsilon$\\\hline
Bravais lattice &Trigonal&Monocl.&Cubic&Cubic&Orthorh.\\
Space group&$R\overline{3}c$ & $C2/m$ &$Fd\overline{3}m$&$Ia\overline{3}$&$Pna2_1$\\
Point group&$\overline{3}m$&$2/m$ &$m\overline{3}m$& $m\overline{3}$&$mm2$\\
Inversion center&$\surd$&$\surd$ &$\surd$&$\surd$&$\times$\\\hline\hline

\end{tabular*}
\end{table}

\begin{figure}
  \includegraphics[width=5.5cm]{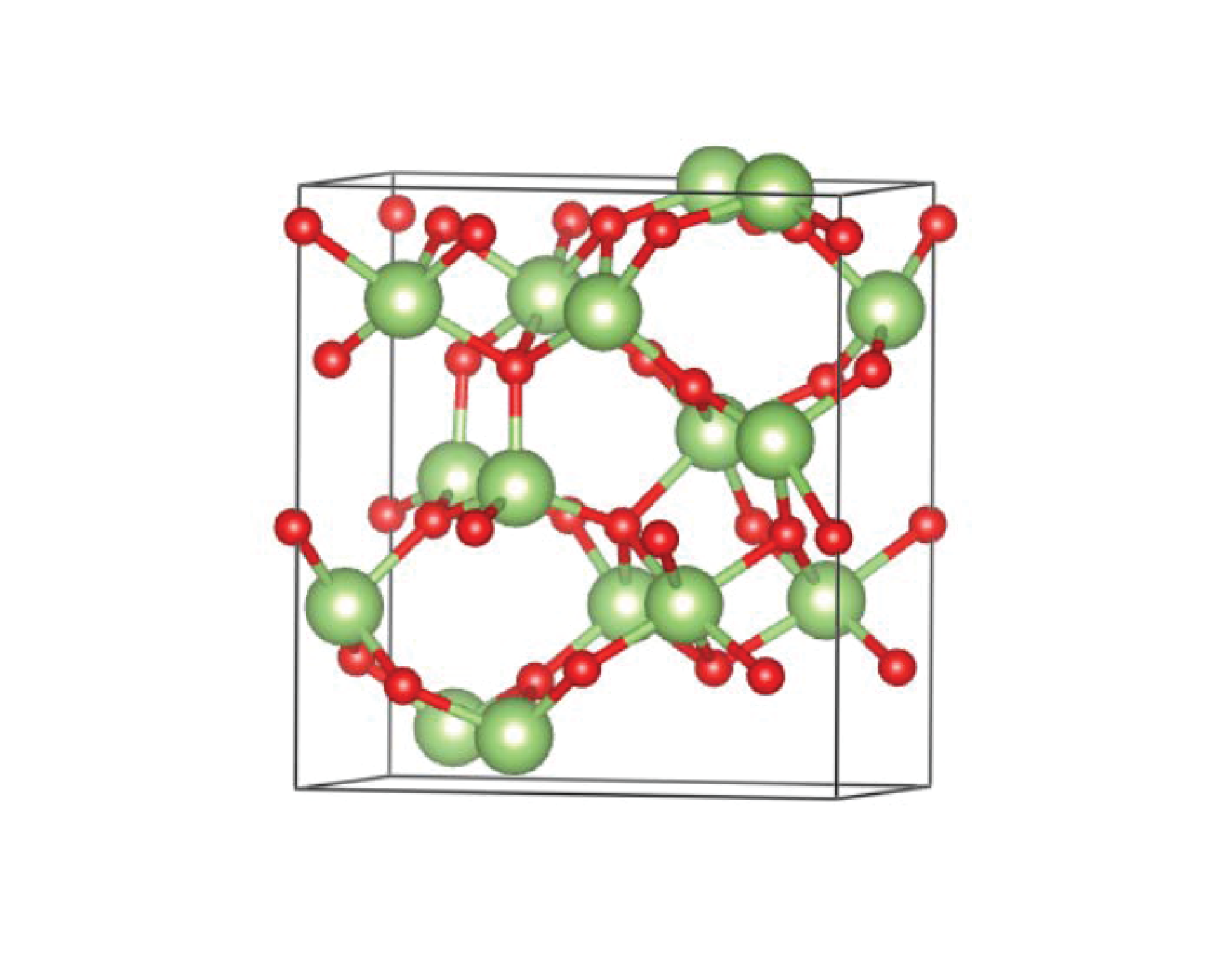}
  \caption{(Color online) The  crystal structure of $\epsilon$-$\mathrm{Ga_2O_3}$. The large green balls represent Ga atoms, and the  small red balls for O atoms. }\label{t0}
\end{figure}

 \section{SYMMETRY ANALYSIS}
 The piezoelectric effect is an electromechanical coupling described by piezoelectric tensors $e_{ijk}$ and $d_{ijk}$, which  are obtained as the sum of ionic and electronic contributions. In
the following, the frequently used Voigt notation is employed, and the mapping of indices is 11$\rightarrow$1,
22$\rightarrow$2, 33$\rightarrow$3, 23$\rightarrow$4, 31$\rightarrow$5 and 12$\rightarrow$6.
The Voigt notation  allows to represent the tensor of elastic constants $C_{ijkl}$, piezoelectric tensors $e_{ijk}$ and $d_{ijk}$
as   6$\times$6, 3$\times$6 and 3$\times$6  matrix $C_{ij}$, $e_{ij}$ and $d_{ij}$, with a maximum of 21, 18 and 18
independent elements. The number of independent
components can be reduced due to  the crystal
symmetry in $C_{ij}$, $e_{ij}$ and $d_{ij}$ tensors. The $\epsilon$-$\mathrm{Ga_2O_3}$ has the $mm2$ point group symmetry, giving:
 \begin{equation}\label{pe1}
 e=
  \left(
    \begin{array}{cccccc}
     0&0&0& 0 & e_{15} & 0 \\
    0&0&0& e_{24} & 0 & 0 \\
      e_{31} & e_{32} & e_{33} & 0&0&0 \\
    \end{array}
  \right)
   \end{equation}

 \begin{equation}\label{pe2}
 C=
  \left(
    \begin{array}{cccccc}
     C_{11}&C_{12}&C_{13}& 0 & 0 & 0 \\
    C_{12}&C_{22}&C_{23}& 0 & 0 & 0 \\
      C_{13}&C_{23}&C_{33}& 0 & 0 & 0 \\
        0&0&0& C_{44} & 0 & 0 \\
    0&0&0& 0 & C_{55} & 0 \\
      0 & 0 &0 & 0&0&C_{66} \\
    \end{array}
  \right)
   \end{equation}

The elastic   tensor $C_{ij}$ and piezoelectric stress tensor $e_{ij}$  can be attained by  density functional theory (DFT) calculations, and the piezoelectric strain tensor $d_{ij}$ can be calculated by the relation:
\begin{equation}\label{pe3}
    e=dC
 \end{equation}
and
 \begin{equation}\label{pe4}
 d=
  \left(
    \begin{array}{cccccc}
     0&0&0& 0 & d_{15} & 0 \\
    0&0&0& d_{24} & 0 & 0 \\
      d_{31} & d_{32} & d_{33} & 0&0&0 \\
    \end{array}
  \right)
   \end{equation}

\section{Electronic structures}
 The crystal structure of  $\epsilon$-$\mathrm{Ga_2O_3}$ has 16 (24) Ga (O) atoms at four (six) different Wyckoff positions
4a, which is plotted in \autoref{t0}. Within the DFT \cite{1}, a full-potential linearized augmented-plane-waves method
is used to investigate electronic structures  of  $\mathrm{Ga_2O_3}$ by using  WIEN2k  code\cite{2}.
 We use Tran and Blaha's mBJ approach for the exchange potential (plus LDA
correlation potential)\cite{mbj},  and the popular GGA of Perdew, Burke and  Ernzerhof  (GGA-PBE)\cite{pbe} to do comparative studies
 We use a 12 $\times$ 7 $\times$ 6 k-point meshes in the first Brillouin zone (BZ) for the self-consistent calculation,  make harmonic expansion up to $\mathrm{l_{max} =10}$ in each of the atomic spheres, and set $\mathrm{R_{mt}*k_{max} = 8}$.

\begin{table}
\centering \caption{The atomic coordinates of $\epsilon$-$\mathrm{Ga_2O_3}$ with $a$=5.06 $\mathrm{{\AA}}$, $b$=8.69 $\mathrm{{\AA}}$ and $c$=9.30 $\mathrm{{\AA}}$. }\label{tab0-1}
  \begin{tabular*}{0.48\textwidth}{@{\extracolsep{\fill}}cccc}
  \hline\hline
atom & x &  y & z\\\hline
Ga1     &   0.18017 &  0.15153  & 0.99762\\
Ga2     &   0.81334 &  0.16181 &  0.30879\\
Ga3     &   0.19165  & 0.15083  & 0.58692\\
Ga4     &  0.67799   &0.03128   &0.79570\\
O1      &    0.97429  & 0.32590 &  0.42764\\
O2       &   0.52161   &0.48778  & 0.43308\\
O3        &  0.65030   &0.00345   &0.20151\\
O4         &0.15460   &0.15917   &0.19757\\
O5       &  0.84997   &0.17145   &0.67252\\
O6        & 0.52301   &0.16682   &0.93836\\\hline\hline
\end{tabular*}
\end{table}

\begin{figure*}
  \includegraphics[width=6.5cm]{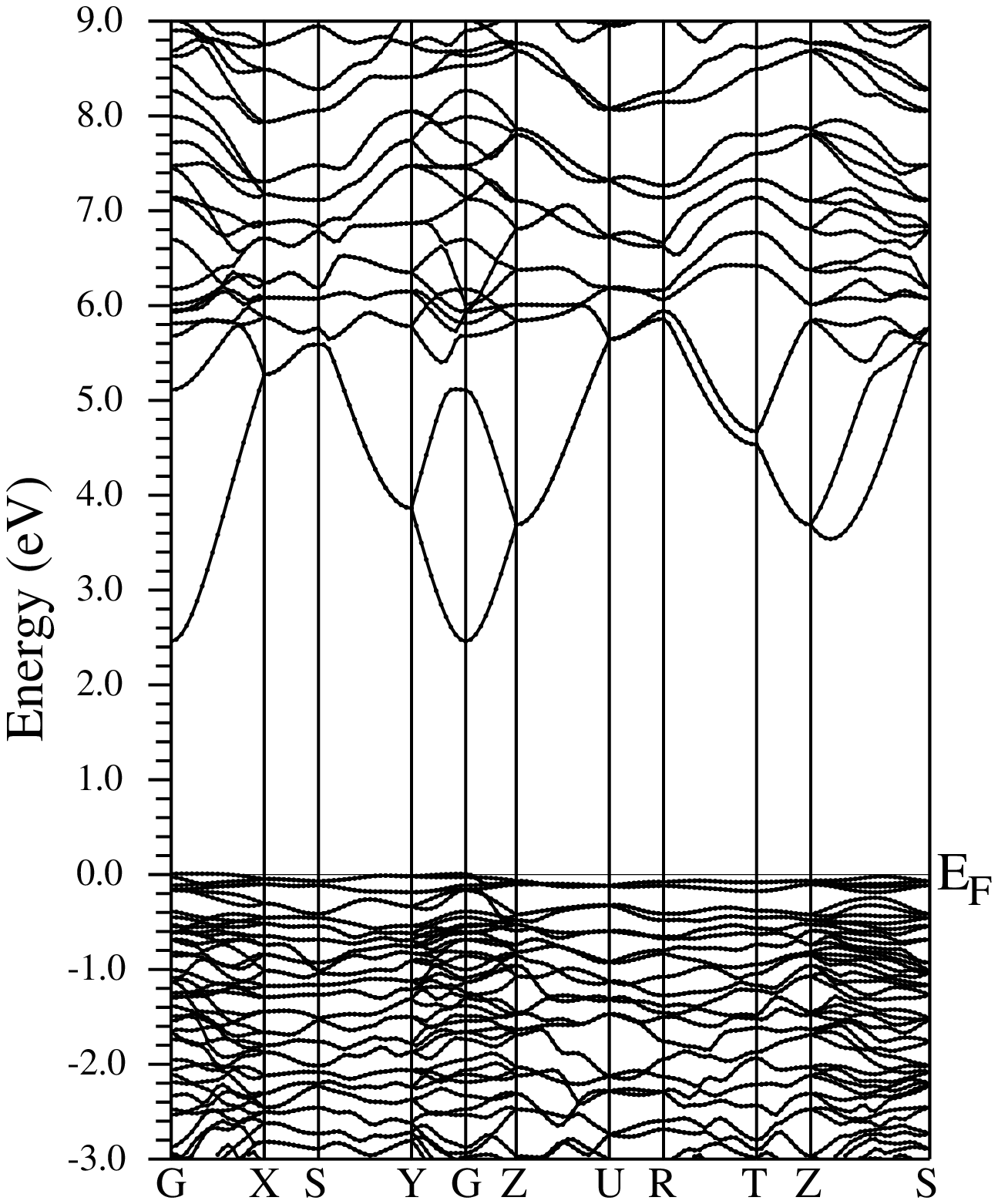}
  \includegraphics[width=6.5cm]{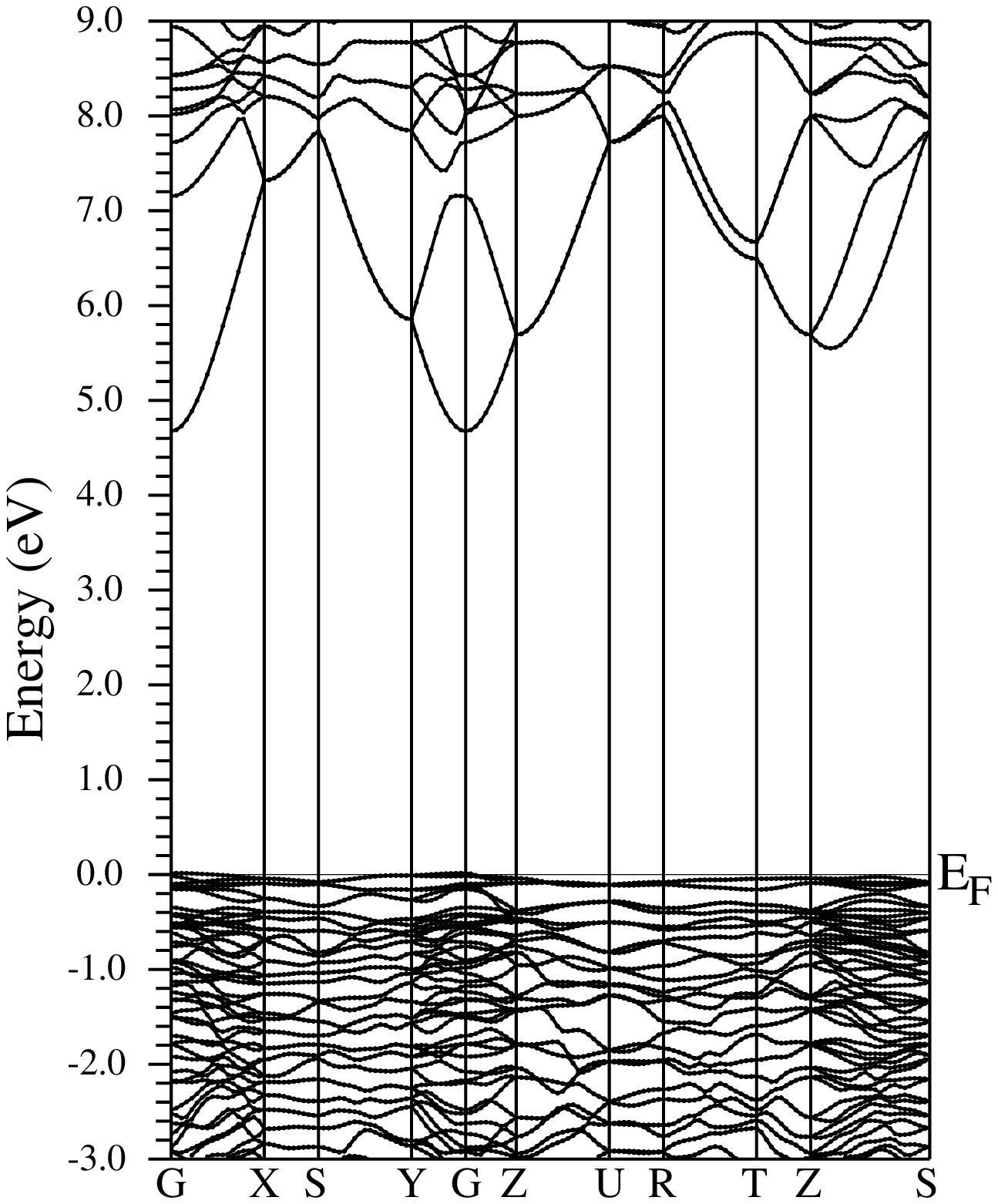}
  \caption{The energy band structures  of   $\epsilon$-$\mathrm{Ga_2O_3}$  using GGA (Left) and mBJ (Right). }\label{t1}
\end{figure*}

The optimized  structure-related data are summarized
in \autoref{tab0-1} by using GGA, which agree well with previous calculated values\cite{g12}. Firstly, the popular GGA is used to perform the self-consistent calculation,  and  the improved
exchange-correlation functional including the mBJ exchange
potential is adopted, which can improve the semiconductor gaps and d state positions for
many kinds of materials.  The energy bands  calculated with GGA and
mBJ are presented in \autoref{t1}.  The GGA gap
value is 2.45 eV, and 4.66 eV for mBJ functional. Our  GGA gap accords with other GGA value 2.465 eV \cite{g12}.
 The mBJ gap is very close to HSE  one (4.26 eV)\cite{g14} and B3LYP one (4.62 eV)\cite{g13}, and  is also close to
experimental value 4.41 eV by ARPES experiments \cite{g14}and 4.6 eV indicated by photoconductivity and optical absorption\cite{g15}.
Both GGA and mBJ results show
a  CBM at the G point.
A quasidirect gap is observed with the
valence band maximum (VBM) is a bit off G in the G-X direction. The energy difference between G point and VBM only is 1.2 meV with GGA and 0.2 meV with mBJ. The experimental data suggest that the VBM is
at or near  the zone centre\cite{g14}.
The experimental gap for  $\beta$-$\mathrm{Ga_2O_3}$  has been reported in the
range of 4.6-4.9 eV\cite{g0,g2}.  The  mBJ  is used to study the electronic structures of  $\beta$-$\mathrm{Ga_2O_3}$ ($a$=12.29 $\mathrm{{\AA}}$, $b$=3.05 $\mathrm{{\AA}}$, $c$=5.81  $\mathrm{{\AA}}$ and $\beta$=103.77). The mBJ gap value of $\beta$-$\mathrm{Ga_2O_3}$ is 4.61 eV, which shows that mBJ
can reproduce the gap of $\mathrm{Ga_2O_3}$  very well.
We also determine the effective mass tensor for electrons at
the CBM, and the resulting results in units of the
free electron mass $m_0$ are: $m_{xx}$=0.237, and $m_{yy}$=$m_{zz}$=0.235, which shows that the anisotropy  is rather small.
These effective masses are very close to ones of monoclinic and rhombohedral cases\cite{g12}.

\begin{table}
\centering \caption{For $\epsilon$-$\mathrm{Ga_2O_3}$, bulk modulus ($B$), Shear modulus ($G$), Young's modulus ($E$), Poisson's ratio ($\nu$), longitudinal wave velocity ($\upsilon_l$), transverse wave velocity ($\upsilon_t$),
average wave velocity ($\upsilon_a$) and Debye temperature ($\Theta_D$). }\label{tab0}
  \begin{tabular*}{0.48\textwidth}{@{\extracolsep{\fill}}cccc}
  \hline\hline
$B$ (GPa)& $G$ (GPa) &  $E$ (GPa) & $\nu$\\\hline
209.22 &82.50&218.75&0.33\\\hline\hline
$\upsilon_l$ (km/s)&$\upsilon_t$ (km/s) & $\upsilon_a$ (km/s)&  $\Theta_D$ (K)\\\hline
7.24&3.68&4.12 &565.81\\\hline\hline
\end{tabular*}
\end{table}
\begin{table}[htb]
\centering \caption{The elastic constants $C_{ij}$  of  ZnO, AlN and GaN, and the unit is GPa.}\label{tab-ela}
  \begin{tabular*}{0.49\textwidth}{@{\extracolsep{\fill}}cccccc}
  \hline\hline
  Name   &$C_{11}$ & $C_{12}$   & $C_{13}$& $C_{33}$&$C_{44}$\\\hline
 ZnO &    204.3&133.2&115.8&209.3&34.3                                                             \\
  AlN &     392.0&141.6&105.5&372.2&112.5                                                               \\
  GaN &     344.8&133.4&93.9&379.6&89.6                                                              \\\hline\hline
\end{tabular*}
\end{table}
\begin{table*}
\centering \caption{Piezoelectric coefficients $e_{ij}(d_{ij})$  of  ZnO, AlN and GaN, and the unit is $\mathrm{C/m^2}$ (pm/V).  The related experimental values of  ZnO\cite{zno}, AlN\cite{aln} and GaN\cite{aln,aln-1,aln-2,aln-3}are shown in  parentheses. }\label{tab-ya}
  \begin{tabular*}{0.98\textwidth}{@{\extracolsep{\fill}}ccccccc}
  \hline\hline
  Name &$e_{33}$ & $e_{31}$   & $e_{15}$& $d_{33}$ &$d_{31}$&$d_{15}$\\\hline
  ZnO &    1.33 (0.96)&    -0.65 (-0.62)& -0.49 (-0.37) & 13.63 (12.3) & -6.59 (-5.1) & -14.20 (-8.3)                                                        \\
  AlN &    1.61 (1.55)&    -0.65 (-0.58)& -0.34 (-0.48) & 5.65 (5.53) & -2.34 (-2.65) & -3.00 (-4.07)                                                            \\
  GaN &   0.67 (1.00\cite{aln})&    -0.37 (-0.36\cite{aln})& -0.23 (-0.30\cite{aln}) & 2.38 (3.1\cite{aln-1}) & -1.24 (-1.0\cite{aln-2}) & -2.52 (-3.1\cite{aln-3})                                                             \\\hline\hline
\end{tabular*}
\end{table*}

\section{Piezoelectric properties}
The elastic   tensor $C_{ij}$ and piezoelectric stress tensor $e_{ij}$ are obtained by using
DFPT\cite{pv6}  as implemented
in VASP code\cite{pv1,pv2,pv3}. The relaxed-ion elastic   tensor  and piezoelectric stress  tensor are obtained from the sum of ionic and electronic
contributions. Within DFPT,  the electronic
and ionic contributions to the piezoelectric tensor can be calculated directly in the VASP code.
A 12$\times$7$\times$6 k-point
mesh is used, and  the exchange-correlation interactions are
treated using the GGA-PBE  with a kinetic-energy cutoff of 450 eV.
It is noted that the order of indices in VASP code is 1(XX), 2(YY), 3(ZZ), 6(XY), 4(YZ), 5(ZX), and we have changed into  the normal order in the following results for elastic and piezoelectric  tensors.
The  elastic  tensor $C_{ij}$ are given (in GPa):
 \begin{equation}\label{pe5}
  \left(
    \begin{array}{cccccc}
     354.77 &   165.45  &   142.20 &      0.  &     0  &     0\\
    165.45  &   316.08  &   150.21  &     0  &     0  &     0\\
    142.20  &   150.21  &   302.61  &     0  &     0   &    0\\
      0  &     0   &    0  &    82.75   &    0   &    0\\
      0   &    0   &    0   &    0 &     61.75   &    0\\
      0    &   0   &    0   &    0   &    0   &  102.61\\
    \end{array}
  \right)
   \end{equation}

Based on $C_{ij}$, average mechanical properties of $\epsilon$-$\mathrm{Ga_2O_3}$ can be attained, including
bulk modulus, Shear modulus, Young's modulus, Poisson's ratio, longitudinal wave velocity, transverse wave velocity,
average wave velocity  and Debye temperature.  The  Born  criteria of mechanical stability for  an orthorhombic crystal is\cite{ela}:
\begin{equation}\label{c}
    C_{11}>0, C_{44}>0, C_{55}>0, C_{66}>0\\
\end{equation}
\begin{equation}\label{c}
    C_{11}C_{22} > C_{12}^2\\
\end{equation}
\begin{equation}\label{c}
    C_{11}C_{22}C_{33}+2C_{12}C_{13}C_{23}-C_{11}C_{23}^2-C_{22}C_{13}^2-C_{33}C_{12}^2 > 0
\end{equation}
The calculated $C_{ij}$ satisfy these conditions, proving that that $\epsilon$-$\mathrm{Ga_2O_3}$ is  mechanically stable.
The related data are summarized in \autoref{tab0}.
The bulk (shear) modulus $B$ ($G$) manifests the resistance
to fracture (plastic deformation).
A high (low) $B/G$ ratio may indicates
its ductility (brittleness), and the critical value
 is around 1.75,  which can be used to separate ductile and brittle materilas.
The value of $\epsilon$-$\mathrm{Ga_2O_3}$ is 2.54, and it can  be classified as a ductility
material.

The piezoelectric stress tensor $e_{ij}$ are shown (in $\mathrm{C/m^2}$):
 \begin{equation}\label{pe6}
  \left(
    \begin{array}{cccccc}
         0     &    0      &   0     &    0   & 0.595   &     0\\
         0      &   0     &    0    &0.194     &    0      &   0\\
    0.011 & -0.319   & 0.941     &    0      &   0     &    0\\
    \end{array}
  \right)
   \end{equation}
The piezoelectric strain tensor $d_{ij}$ are derived by  \autoref{pe3}, giving (in pm/V):
 \begin{equation}\label{pe7}
  \left(
    \begin{array}{cccccc}
         0     &    0      &   0     &    0   & 9.622   &     0\\
         0      &   0     &    0    &2.345     &    0      &   0\\
    -0.489 & -3.060  & 4.858     &    0      &   0     &    0\\
    \end{array}
  \right)
   \end{equation}
The $\epsilon$-$\mathrm{Ga_2O_3}$ possesses five independent
components of the piezoelectric tensor, namely $d_{31}$,  $d_{32}$, $d_{33}$,  $d_{15}$ and  $d_{24}$.
The magnitudes of $d_{ij}$ range from 0.489 pm/V to 9.622 pm/V, the $d_{33}$ and  $d_{15}$ of  which are
comparable and even higher than commonly used
 piezoelectric materials such as $\alpha$-quartz, ZnO, AlN  and GaN\cite{zno,aln,aln-1,aln-2,aln-3,aln-4}.
We note that $d_{31}$ is smaller by 1 order  of magnitude compared to other $d_{ij}$, which is due to very small $e_{31}$.
The $d_{15}$ is the largest among the $d_{ij}$, which is due to the large $e_{15}$ and the smallest $C_{55}$ ($d_{15}$=$e_{15}$/$C_{55}$).
Our calculated $d_{ij}$ are close to previous theoretical values ($d_{32}$=-3.43 pm/V, $d_{33}$=4.06 pm/V, $d_{24}$=2.69 pm/V, $d_{15}$=14.60 pm/V) except $d_{31}$ (1.37 pm/V)\cite{g16}.  In previous calculations, the  elastic  tensor $C_{ij}$ are attained by fitting the DFT-calculated
unit-cell energy  to a series of  strain states, and the piezoelectric stress tensor $e_{ij}$ are calculated  by evaluating the change of
unit-cell polarization after imposing small strain\cite{g16}. Here, these tensors are calculated by DFPT. To ensure the reliability of our results or method, the piezoelectric  properties of ZnO, AlN  and GaN  with $P6_3mc$ space group are also studied by DFPT. Due to $6mm$ point group of $P6_3mc$, they  have five independent elastic constants ($C_{11}$, $C_{12}$, $C_{13}$, $C_{33}$ and $C_{44}$), and three piezoelectric  constants ($e/d_{31}$, $e/d_{33}$ and $e/d_{15}$).  The elastic constants  of ZnO, AlN  and GaN are shown in \autoref{tab-ela}, which agree well with previous calculated values\cite{ela1}.
The piezoelectric  tensors of ZnO, AlN  and GaN  are summarized in \autoref{tab-ya}, along with the related experimental values of  ZnO\cite{zno}, AlN\cite{aln} and GaN\cite{aln,aln-1,aln-2,aln-3}. Our calculated piezoelectric  tensors of ZnO, AlN  and GaN are in reasonable agreement
with experiments. The deviation may be because  the related experiments
are carried out on constrained epitaxial samples. Thus, our predicted piezoelectric  tensors of $\epsilon$-$\mathrm{Ga_2O_3}$ should be receivable.

\section{Discussions and Conclusion}
It is clear that  mBJ gives much better energy gap of $\epsilon$-$\mathrm{Ga_2O_3}$ than
GGA  toward the experimental values. Although HSE or B3LYP also can give reasonable energy gap, they need more CPU
time and memory than mBJ. Thus, mBJ may be more  suitable for dopant studies of $\epsilon$-$\mathrm{Ga_2O_3}$.
It is noted that the energy gap and the  effective masses (at CBM) of $\epsilon$-$\mathrm{Ga_2O_3}$ are very close to ones of $\beta$ case,
but $\epsilon$ phase  shows good piezoelectric  properties, which can add  more freedom for electronic devices.

In summary, the electronic structures have been studied by GGA and mBJ, and the elastic and piezoelectric tensors are attained by DFPT.
The mBJ gap is consistent with previously calculated HSE  or B3LYP  one, and has very better agreement with experiment than GGA one.
The values of $d_{ij}$ are found to be
comparable to or even superior than conventional
piezoelectric materials such as $\alpha$-quartz, ZnO, AlN and GaN.
These results show the possibility of employing piezoelectric effects  in $\epsilon$-$\mathrm{Ga_2O_3}$ for electronics and energy applications.
Our works can stimulate further experimental works to study piezoelectric  properties of  $\epsilon$-$\mathrm{Ga_2O_3}$.

\begin{acknowledgments}
This work is supported by the Natural Science Foundation of Shaanxi Provincial Department of Education (19JK0809). We are grateful to the Advanced Analysis and Computation Center of China University of Mining and Technology (CUMT) for the award of CPU hours and WIEN2k/VASP software to accomplish this work.
\end{acknowledgments}

\end{document}